\documentclass[journal=jpcbfk,manuscript=article,layout=traditional]{achemso}

\usepackage[version=3]{mhchem} 
\usepackage[T1]{fontenc}       
\usepackage{xspace}
\usepackage{mathtools}

\newcommand{\DISS}{\omega}
\newcommand{\TOT}{\DISS_{\text{tot}}}
\newcommand{\PTOT}{P_{\text{tot}}}

\newcommand{\RCF}{k_{ij}^+}
\newcommand{\RCR}{k_{ij}^-}
\newcommand{\RCFONETWO}{k_{12}^+}
\newcommand{\RCFTWOONE}{k_{21}^+}
\newcommand{\RCFTWOTHREE}{k_{23}^+}
\newcommand{\RCFTHREEONE}{k_{31}^+}
\newcommand{\RCRONETWO}{k_{12}^-}
\newcommand{\RCRTWOONE}{k_{21}^-}
\newcommand{\RCRTWOTHREE}{k_{23}^-}
\newcommand{\RCRTHREEONE}{k_{31}^-}

\newcommand{\DOPT}{\DISS^*}

\newcommand{\DELD}{\Delta\DISS}
\newcommand{\BARE}{k_{ij}^0}

\newcommand{\BAREONETWO}{k_{12}^0}
\newcommand{\BARETWOONE}{k_{21}^0}
\newcommand{\BARETWOTHREE}{k_{23}^0}
\newcommand{\BARETHREEONE}{k_{31}^0}
\newcommand{\MD}{\text{d}}
\newcommand{\forwardlabile}{forward labile\xspace}
\newcommand{\reverselabile}{reverse labile\xspace}

\newcommand{\FEC}{free energy component\xspace}
\newcommand{\FECS}{free energy components\xspace}
\newcommand{\FECSCAP}{Free energy components\xspace}

\newcommand{\FEB}{free energy budget\xspace}

\newcommand{\DISSONE}{\DISS^{\text{I}}}
\newcommand{\DISSTWO}{\DISS^{\text{II}}}
\newcommand{\TOTONE}{\DISS^{\text{I}}_{\text{tot}}}
\newcommand{\TOTTWO}{\DISS^{\text{II}}_{\text{tot}}}
\newcommand{\DISSONEOPT}{\DISS^{\text{I*}}}
\newcommand{\DISSTWOOPT}{\DISS^{\text{II*}}}
\newcommand{\DOPTONELOAD}{\DISS_{ij}^{\text{I}*,\text{load}}}
\newcommand{\DOPTONENOLOAD}{\DISS_{ij}^{\text{I}*,\text{no load}}}
\newcommand{\SPLITONE}{\delta^{\text{I}}}
\newcommand{\SPLITTWO}{\delta^{\text{II}}}
\author{Aidan I. Brown}
\email{aidanb@sfu.ca}
\author{David A. Sivak}
\email{dsivak@sfu.ca}
\affiliation[Simon Fraser University]
{Department of Physics, Simon Fraser University, Burnaby, BC, V5A1S6 Canada}

\let\oldnewpage=\newpage

\let\newpage\relax\title[Allocating and splitting free energy to maximize molecular machine flux]
  {Allocating and splitting free energy to maximize molecular machine flux}

\begin{document}


%
%
%
%
%

\begin{abstract}
Biomolecular machines transduce between different forms of energy.  These machines make directed progress and increase their speed
by consuming free energy, typically in the form of nonequilibrium chemical concentrations. 
Machine dynamics are often modeled by transitions between a set of discrete metastable conformational states. In general, the free energy change associated with each transition can increase the forward rate constant, decrease the reverse rate constant, or both. In contrast to previous optimizations,
we find that in general flux is neither maximized by devoting all free energy changes to increasing forward rate constants nor by solely decreasing reverse rate constants.  Instead the optimal free energy splitting depends on the detailed dynamics.
Extending our analysis to machines with vulnerable states (from which they can break down), in the strong driving corresponding to \emph{in vivo} cellular conditions, processivity is maximized by reducing the occupation of the vulnerable state.
\end{abstract}


\section{Introduction}

Molecular machines such as kinesin~\cite{clancy11}, helicase~\cite{caruthers02}, and ubiquitin ligase~\cite{cardozo04} perform diverse tasks inside cells. These machines typically convert nonequilibrium chemical concentrations, maintained by other machinery in the cell~\cite{boyer97,fernie04}, into directed motion or work~\cite{vale00,bustamante01}. These microscopic machines operate stochastically~\cite{astumian94}, with their fluctuating progress now experimentally observable with improving resolution (see \emph{e.g.}\ Isojima \emph{et al}~\cite{isojima16}).

Quantitative models of molecular machines are pervasive, \emph{e.g.}\ to investigate forces~\cite{fisher99}, efficiency~\cite{schmiedl08}, and performance of different driving mechanisms~\cite{wagoner16,astumian15}. Each model usually treats molecular machine dynamics as a set of transitions between discrete metastable states~\cite{thomas01}, as diffusion on a continuous energy landscape~\cite{astumian16}, or a combination of the two~\cite{xing05}.

In this paper we study how the flux of cyclic machines, a posited driver of evolutionary fitness~\cite{nguyen17}, is affected by the details of free energy changes over a set of discrete states.

Earlier models considered how the influence of a load is quantitatively split between forward and reverse transitions~\cite{wagoner16,schmiedl08,thomas01,fisher99}.
In this work we find that by extending the splitting to all \FECS, not just loads, there is no single splitting scheme that always maximizes the flux.
Additionally, optimizing the allocation of a single \FEC compensates for any sub-optimal fixed allocations of other \FECS.

We also examine how to maximize progress for transiently processive cyclic machines with vulnerable states, from which the machine can break down or `escape' (\emph{e.g.}\ kinesin can dissociate from a microtubule~\cite{milic14}). For strong driving, free energy is simply allocated to reduce the occupation of the state vulnerable to escape. However, under specific conditions, the optimal allocation is reversed, counterintuitively increasing processivity by increasing occupation of the vulnerable state.

\section{Model}

\subsection{Discrete states}

\begin{figure}[tbp] 
	\centering
	\hspace{-0.100in}\includegraphics[width=3.35in]{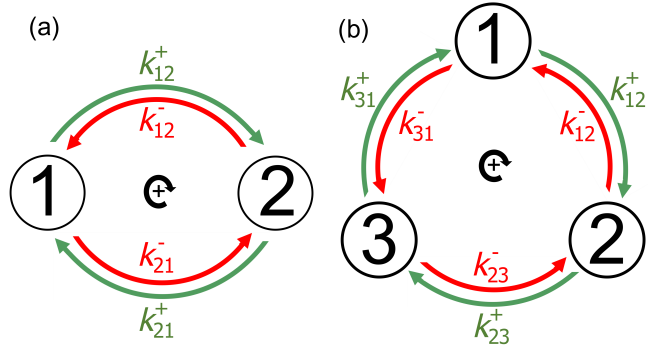}\\
	\caption{\label{fig:Diagrams} 
{\bf Discrete-state cycles with (a) two and (b) three states.} Transitions occur in both the forward (green arrows) and reverse (red arrows) directions for each pathway between two states, described by respective rate constants $\RCF$ and $\RCR$.
}
\end{figure}

We consider cycles with two or three states (Fig.~\ref{fig:Diagrams}), often used for models of molecular motors or other driven systems inside living cells~\cite{kolomeisky03,qian06}. 
For our model cycles, each forward rate constant $\RCF$ describes transitions from state $i$ to state $j$, with a corresponding reverse rate constant $\RCR$ for transitions from state $j$ to state $i$. To preserve microscopic reversibility~\cite{astumian15, fisher99}, any transition with a nonzero forward rate cannot be irreversible, and must also have a nonzero reverse rate.

Forward transitions from state $i$ to state $j$ occur at rate $\RCF P_i$, for probability $P_i$ in state $i$. Reverse transitions from state $j$ to state $i$ occur at reverse rate $\RCR P_j$. (The two-state cycle has two pathways, 12 and 21, each with a forward and reverse direction, and each representing distinguishable physical transition mechanisms.)

To preferentially drive such cycles in a particular direction, cells use nonequilibrium concentrations of reacting chemical species, such as adenosine triphosphate (ATP), adenosine diphosphate (ADP), and inorganic phosphate ($\text{P}_{\text{i}}$)~\cite{phillips12}. ATP hydrolysis into ADP and $\text{P}_{\text{i}}$ yields a free energy $\Delta G$ that depends on the respective concentrations~\cite{phillips12},
\begin{equation}
\Delta G = \Delta G_0 + k_{\text{B}}T\ln\frac{[\text{ADP}][\text{P}_{\text{i}}]}{[\text{ATP}]} \ .
\end{equation}
Here $\Delta G_0 \equiv -k_{\text{B}}T \ln\left([\text{ADP}]_{\text{eq}}[\text{P}_{\text{i}}]_{\text{eq}}/[\text{ATP}]_{\text{eq}}\right)$, $k_{\text{B}}$ is Boltzmann's constant, and $T$ is the temperature of the thermal bath exchanging heat with the system. ATP hydrolysis provides free energy $|\Delta G|\sim 20 k_{\text{B}}T$ under typical physiological conditions~\cite{phillips12}. From here on we set $k_{\text{B}}T = 1$, thus measuring all free energies in units of the thermal energy scale $k_{\text{B}}T$.

The free energy allocation $\DISS_{ij}$ of a given transition path fixes the ratio between the forward and reverse rate constants~\cite{wagoner16,pietzonka16,rao16},
\begin{equation}
\label{eq:ratio}
\DISS_{ij} = \ln \frac{\RCF}{\RCR} \ . 
\end{equation}
Without $\DISS_{ij}$ to provide a bias, the full forward and reverse rate constants $k_{ij}^{+/-}$ each equal the `bare' rate constants $\BARE$. $\DISS_{ij} = -\Delta G_{ij}$, where $\Delta G_{ij}$ is the free energy change of the machine and its surroundings over the transition. $\Delta G$ over a given cycle must be negative to drive net progress in the forward direction (on average).

\subsection{Biasing rates with free energy}

$\DISS_{ij}$ can be composed of several different \FECS, including but not limited to conformational changes, binding/unbinding, or a change in potential.

The free energy change $\Delta G^{\text{mach}}_{ij}$ of the molecular machine over a transition includes free energy differences between distinct machine conformations, as well as changes in the free energy of any molecules bound to the machine (\emph{e.g.}\ hydrolysis of ATP to ADP and $\text{P}_{\text{i}}$), as the free energy changes of the machine and bound species cannot be separated~\cite{hill81,hill83}.

$\Delta G^{\text{bind}}_{ij} = -\mu$ is the free energy change of the solution when a molecule binds the machine, leaving the solution, with $\mu$ the chemical potential of the molecule given the concentration in solution. $\Delta G^{\text{bind}}_{ij}$ includes molecules typically considered `fuel', such as ATP; and those typically considered cargo. For a molecule that unbinds from the machine and joins the solution, $\Delta G^{\text{unbind}}_{ij} = \mu$.

$\Delta G^{\text{pot}}_{ij}$ represents the free energy change of the machine, or any object attached to the machine, due to an external potential or gradient, \emph{e.g.}\ due to a bead attached to kinesin that is also in an optical trap.

We confine our attention in this paper to `reversible' \FECS, whereby any free energy expended in a forward reaction is recovered by the reverse reaction. Thus we exclude from consideration any omnidirectional free energy dissipation, such as that due to friction when pulling a load through a viscous medium.  

Excepting $\Delta G^{\text{mach}}_{ij}$, we expect that the \FECS are relatively fixed (\emph{i.e.}\ unable to be changed by a modification to the machine). $\sum\Delta G^{\text{mach}}_{ij} = 0$, as the machine returns to the same state after completing each cycle -- abiding by this constraint, variation of $\Delta G^{\text{mach}}_{ij}$ (nominally through machine mutations) can be used to optimize machine operation.

We model each \FEC $\DISS_{ij}^k = -\Delta G^k_{ij}$ with a different splitting factor~\cite{schmiedl08,wagoner16} $\delta_{ij}^k\in[0,1]$, which describes how the effect of $\DISS_{ij}^k$ is divided between forward and reverse rate constants,
\begin{equation}
\label{eq:RateSchemeLoad}
\RCF \propto e^{\sum_k \delta_{ij}^k\DISS_{ij}^k} \ \ \text{and} \ \  \RCR \propto e^{-\sum_k (1-\delta_{ij}^k)\DISS_{ij}^k} \ .
\end{equation}

Earlier studies~\cite{wagoner16,schmiedl08,thomas01,fisher99} considered a molecular motor pulling against a constant opposing force $F$, with each cycle of the motor stepping forward a distance $d$. The transition completing the step transduces free energy $w=Fd$ to the motor's position in the potential. In our framework this corresponds to $\DISS_{ij}^{\text{pot}} = -w$. For steps against this constant force, $\delta_{ij}^{\text{pot}} = 0$ is known as a \emph{power stroke} (PS) and $\delta_{ij}^{\text{pot}} = 1$ is known as a \emph{Brownian ratchet} (BR)~\cite{wagoner16}.

In this paper we primarily consider the case where splitting factors $\delta_{ij}^k$ have the same value $\delta$ for all \FECS. 
This simple framework assumes the transition state is similarly arrayed between the reactant and product along different coordinate axes.
We label $\delta=1$ as \emph{forward labile} (FL) and $\delta = 0$ as \emph{reverse labile} (RL). FL is then similar to BR, and RL to PS~\cite{brown17}.
The choice between PS and BR, or between FL and RL, affects machine performance characteristics~\cite{wagoner16,brown17}, and below we determine how to vary splitting factors to maximize flux.

\section{Maximizing flux}
\label{sec:load}

\subsection{Optimal splitting factors}

Setting all splitting factors $\delta_{ij}^k$ to the same value $\delta$ allows the combination of all $\delta_{ij}^k\DISS_{ij}^k$ terms in Eqs.~\ref{eq:RateSchemeLoad} into single terms $\delta\DISS_{ij}$ and $(1-\delta)\DISS_{ij}$, giving rate constants
\begin{subequations}
\label{eq:RateSchemeGeneral2}
\begin{align}
\RCF &= \BARE e^{\delta \DISS_{ij}} \ , \\
\RCR &= \BARE e^{-(1-\delta)\DISS_{ij}} \ .
\end{align}
\end{subequations}
We consider a two-state cycle, with steady-state flux (hereafter simply flux)~\cite{hill77}
\begin{equation}
\label{eq:TwoStateFlux}
J = \frac{\RCFONETWO\RCFTWOONE - \RCRONETWO\RCRTWOONE}{\RCFONETWO + \RCRONETWO + \RCFTWOONE + \RCRTWOONE} \ .
\end{equation}
Inserting Eq.~\ref{eq:RateSchemeGeneral2} into Eq.~\ref{eq:TwoStateFlux} and differentiating with respect to $\delta$ gives
\begin{align}
\frac{\partial J}{\partial \delta} =& \dfrac{(1 - e^{-\TOT})\left[\splitdfrac{\BAREONETWO \DISS_{21}e^{-\delta \DISS_{21}}(1 + e^{-\DISS_{12}}) +}{\BARETWOONE \DISS_{12}e^{-\delta \DISS_{12}}(1 + e^{-\DISS_{21}})}\right]}{\left[\splitdfrac{\left(\BARETWOONE\right)^{-1} e^{-\delta \DISS_{21}}(1 + e^{-\DISS_{12}}) +} {\left(\BAREONETWO\right)^{-1} e^{-\delta \DISS_{12}}(1 + e^{-\DISS_{21}})}\right]^2} \ .
\end{align}

The cycle proceeds forward when $\TOT = \DISS_{12} + \DISS_{21} > 0$. $\partial J/\partial \delta > 0$ for all $\delta$ when both $\DISS_{12}>0$ and $\DISS_{21} > 0$; for these conditions, $\delta = 1$ (FL) maximizes the flux. For $\delta = 0$ (RL) to maximize the flux requires
\begin{equation}
\BAREONETWO \DISS_{21}(1 + e^{-\DISS_{12}})e^{-\delta \DISS_{21}} + \BARETWOONE \DISS_{12}(1 + e^{-\DISS_{21}})e^{-\delta \DISS_{12}} < 0 \ ,
\end{equation}
a more complicated condition to fulfill, as $\DISS_{12}<0$ and $\DISS_{21} < 0$ cannot be simultaneously fulfilled with $\TOT > 0$. 
Flux can also be maximized for intermediate $\delta$, with $\partial J/\partial\delta$ changing sign (maximizing flux) when
\begin{equation}
\delta = \frac{1}{\DISS_{21} - \DISS_{12}}\ln\left[-\frac{\BAREONETWO \DISS_{21}(1 + e^{-\DISS_{12}})}{\BARETWOONE \DISS_{12}(1 + e^{-\DISS_{21}})}\right] \ .
\end{equation}
The flux-maximizing value of splitting factor $\delta$ thus depends on free energy allocation $\DISS_{ij}$ and bare rate constants $\BARE$.

Although here we set $\delta_{ij}^k = \delta$, in Supplemental Information (SI): Varying splitting factors, we consider splitting factors $\delta_{ij}^k$ that vary across different \FECS and different transitions. Notably, for splitting factors specific to each \FEC $k$ and transition $ij$, if $\DISS_{ij}^k < 0$ then $\delta_{ij}^k = 0$ maximizes the flux, and if $\DISS_{ij}^k > 0$ then $\delta_{ij}^k = 1$ maximizes the flux.

For a molecular motor pulling against a constant force, a forward step has \FEC $\DISS_{ij}^{\text{pot}} = -w < 0$.  
For independent variation of $\delta_{ij}^k$, flux is maximized for $\delta_{ij}^{\text{pot}} = 0$. In this scenario, the optimal splitting factor agrees with Wagoner and Dill's finding that $\delta = 0$ maximizes the power~\cite{wagoner16}. Our results are generally distinct, and do not find a universal optimal $\delta$ value, because we generalize to cycles with multiple states and splitting factors that apply to all \FECS, not just the \FEC associated with pulling against a constant force.

\begin{figure}[tbp] 
	\centering
	\hspace{-0.0in}
	\begin{tabular}{c}
		\hspace{-0.200in}\includegraphics[width=3.5in]{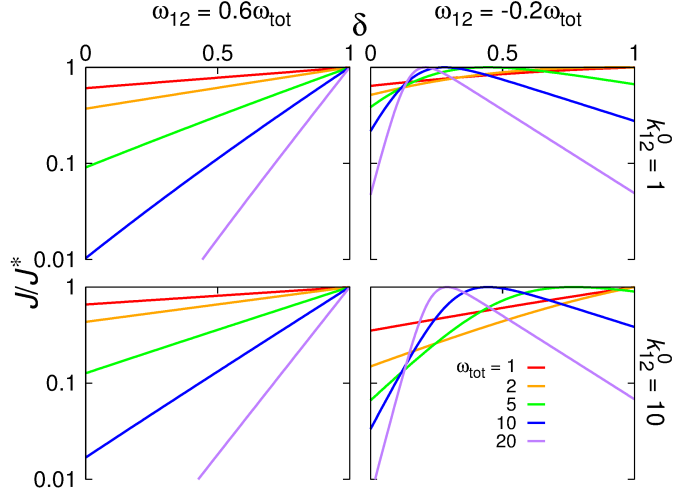}
	\end{tabular}
	\caption{\label{fig:TwoStateSensitivityDelta} 
{\bf Two-state flux sensitivity to splitting factor.} 
Flux ratio $J/J^*$ as a function of the splitting factor $\delta$, for a two-state cycle with bare rate $k_{21}^0 = 1$. $k_{12}^0$, $\TOT$, and $\DISS_{12}$ vary as indicated. $J$ is given by substituting the rate constants of Eq.~\ref{eq:RateSchemeGeneral2} into Eq.~\ref{eq:TwoStateFlux}. $J^*$ is determined in the range $\delta\in[0,1]$.
Optimal flux $J^*$ is specific to each curve on each subplot.
}  
\end{figure}

Fig.~\ref{fig:TwoStateSensitivityDelta} shows the variation in flux as the splitting factor $\delta$ is varied from the flux-maximizing value in the range $\delta\in[0,1]$. The flux can be maximized at an extreme splitting factor value, $\delta = 1$ (left panels) or $0$ (not shown, but possible), or between the extreme values, $\delta\in(0,1)$ (right panels). 
The flux can decrease by more than an order of magnitude away from the optimal $\delta$ value, and decreases faster for larger $\TOT$.

\subsection{Optimal free energy allocation}

Considering Eq.~\ref{eq:RateSchemeGeneral2}, we vary the free energy $\DISS_{ij}$ allocated to each transition to adjust the rate constants and maximize the flux.

For the total \FEB per cycle $\TOT = \DISS_{12} + \DISS_{21}$, the flux-maximizing free energy allocation $\DOPT_{12}$ satisfies (see SI: Varying free energy)
\begin{align}
\label{eq:OptimalDissipation}
\BAREONETWO e^{\delta \DISS_{12}^*}&\left[\delta - (1-\delta)e^{-\DISS_{12}^*}\right] = \\
&\BARETWOONE e^{\delta(\TOT - \DISS_{12}^*)}\left[\delta - (1-\delta)e^{-(\TOT - \DISS_{12}^*)}\right] \ . \nonumber 
\end{align}
This cannot generally be solved for $\DISS_{12}^*$.

For equal splitting factors $\delta_{ij}^k = \delta$, rate constants are determined by the sum of all \FECS $\sum_k\DISS_{ij}^k$, such that if all but one \FEC is fixed, the remaining \FEC can be varied to achieve any desired sum.
This makes maximal flux attainable by only adjusting the free energy allocation of the molecular machine, $\Delta G^{\text{mach}}_{ij}$.

For $\delta = 0$ and $1$ the rate constants reduce to those of our previous work~\cite{brown17}, from which the flux-maximizing free energy allocation $\DISS_{ij}^*$ can be determined. For $\delta=1$ (FL),
\begin{equation}
\label{eq:SimpleOpt1}
\DISS_{12}^* = \frac{1}{2}\TOT + \frac{1}{2}\ln\frac{\BARETWOONE}{\BAREONETWO} \ .
\end{equation}
For $\delta = 0$ (RL), 
\begin{equation}
\label{eq:SimpleOpt2}
\DISS_{12}^* = \frac{1}{2}\TOT - \frac{1}{2}\ln\frac{\BARETWOONE}{\BAREONETWO} \ .
\end{equation}

$\DISS_{ij}$ may be decomposed into multiple \FECS; 
for conceptual clarity we limit our discussion to
two, $\DISS_{ij} = \DISSONE_{ij} + \DISSTWO_{ij}$, but our results trivially generalize.
With $\DOPT_{ij} = \DISSONEOPT_{ij} + \DISSTWOOPT_{ij}$, $\sum\DISSONE_{ij} = \TOTONE$, and $\sum\DISSTWO_{ij} = \TOTTWO$,
then for $\delta = 1$, Eq.~\ref{eq:SimpleOpt1} becomes
\begin{equation}
\label{eq:SimpleOpt3}
\DISSONEOPT_{12} + \DISSTWOOPT_{12} = \tfrac{1}{2}\left(\TOTONE + \TOTTWO\right) + \frac{1}{2}\ln\frac{\BARETWOONE}{\BAREONETWO} \ .
\end{equation}
If one component
is fixed (without loss of generality $\DISSTWO_{ij}$), then $\DISSONE_{ij}$ can vary to preserve the flux-maximizing allocation $\DOPT_{ij} = \DISSONEOPT_{ij} + \DISSTWOOPT_{ij}$. Effectively, the variable component of $\DISS_{ij}$ can compensate for the fixed component to maximize the flux.

In earlier work~\cite{brown17}, we showed that for a range of several $k_{\text{B}}T$ around the optimal allocation, the flux can decrease by more than an order of magnitude. The free energy allocation can thus meaningfully alter the cycle output.

In this section we considered equal splitting factors for all \FECS and transitions: $\delta_{ij}^k = \delta$. In SI: Varying free energy, we consider the more general case where the splitting factor varies over \FECS and transitions, similarly finding that variable \FECS compensate for fixed components.

\subsection{Robustness to variable load}

\begin{figure}[tbp] 
 	\centering
 	\hspace{-0.0in}
 	\begin{tabular}{c}
 		\hspace{-0.200in}\includegraphics[width=3.375in]{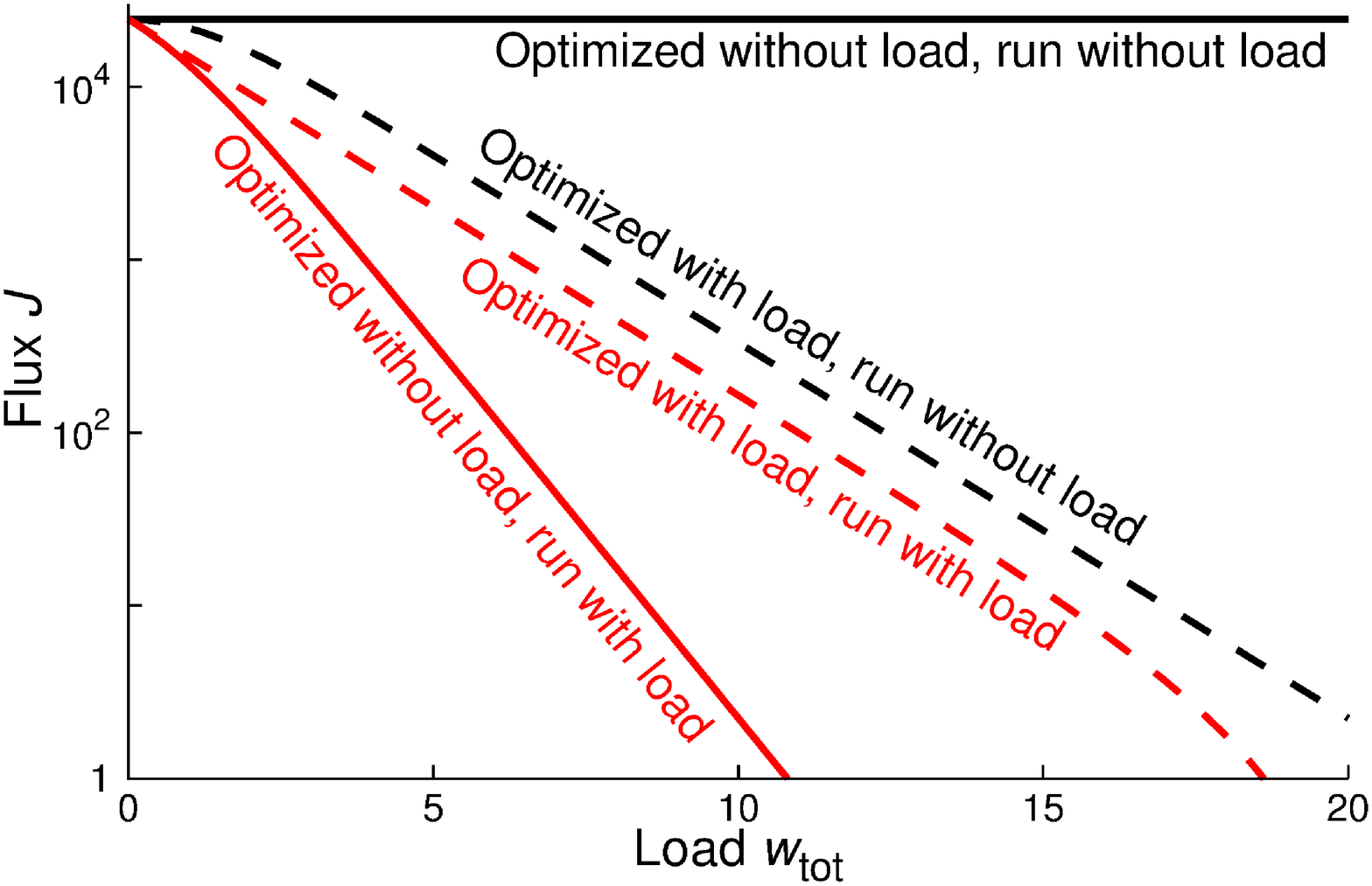}\\
 	\end{tabular}
 	\caption{\label{fig:Forces}   
 {\bf Robustness to variable load.} Flux $J$ vs.\ load $w_{\text{tot}}$ for a two-state \forwardlabile ($\delta_{ij}^k = \delta = 1$) cycle. The load is entirely on the first transition, $\DISSTWO_{12} = -w_{\text{tot}}$, and $\DISSTWO_{21} = 0$. Bare rate constants $\BAREONETWO = 5$, $\BARETWOONE = 1$, and load-free free energy budget $\TOTONE = 20$. Solid curves are for $\DOPTONENOLOAD$ optimized for no load, while dashed curves are for $\DOPTONELOAD$ optimized with load $w_{\text{tot}}$. Black curves show flux without load ($w_{\text{tot}} = 0$), while red curves show flux with load $w_{\text{tot}}$ as indicated by the horizontal axis.}
\end{figure}

The \emph{in vivo} operating environment of molecular machines is diverse, with variable cargo, molecular concentrations, and other factors causing some free energy components to vary over the life-cycle of a machine.
We consider a scenario with two \FECS over a cycle, $\TOTONE$ and $\TOTTWO$. $\TOTONE$ has some fixed value, but $\TOTTWO$ is a load that can be applied or removed, with $\TOTTWO = 0$ (no load) or $\TOTTWO = -w_{\text{tot}} < 0$ (load).
The allocation of $\TOTTWO$ to the various $\DISSTWO_{ij}$ is fixed.
If the allocation of $\DISSONE_{ij}$ is optimized without (with) an applied load, but then the load is applied (removed), the flux will generally be lower than if $\DISSONE_{ij}$ is optimized under the correct conditions. Fig.~\ref{fig:Forces} shows the flux as a function of load for $\TOTONE = 20 k_{\rm B}T$ (that of ATP at physiological conditions) for the four possible scenarios, optimized with ($\DOPTONELOAD$) or without load ($\DOPTONENOLOAD$) and subsequently applying or not applying a load. 

The fluxes for the four scenarios have a fixed order, independent of model parameters. Without load, cycle flux is higher under the correctly optimized allocation $\DOPTONENOLOAD$ than under $\DOPTONELOAD$ incorrectly optimized for load; under $\DOPTONELOAD$ the flux is higher without load than with, because adding a resistive load always reduces the flux; and with load, flux under the correctly optimized $\DOPTONELOAD$ is higher than under the incorrectly optimized $\DOPTONENOLOAD$.
Due to this fixed ordering, the flux changes less when the load is applied or removed if the \FEC allocation is optimized with a load ($\DOPTONELOAD$), compared to if the allocation is optimized for no load ($\DOPTONENOLOAD$).

\section{Processivity}
\label{sec:processivity}

Many cycles are transiently processive, eventually `escaping' from their processive mode of operation, precluding further progress. For example, a transport motor can detach from its cytoskeletal track and diffuse away, effectively ending its forward transport~\cite{milic14,hodges07}.

We model escape from a single `vulnerable' state (without loss of generality, state 2), consistent with experiments suggesting kinesin primarily detaches from a subset of states~\cite{milic14,andreasson15,muthukrishnan09,nam15,toprak09}. The vulnerable state has an escape rate constant $k_{\text{esc}}$. The states which are not vulnerable have the same dynamics as earlier. For the two-state cycle, escape produces modified state 2 dynamics
\begin{equation}
\label{eq:ShowEscapeMain}
\frac{\MD P_2}{\MD t} = (\RCFONETWO + \RCRTWOONE)P_1 - (\RCRONETWO + \RCFTWOONE)P_2 - k_{\text{esc}}P_2 \ .
\end{equation}
For $k_{\text{esc}} > 0$, probability leaves the cycle and there are no steady states for occupation probabilities $P_i$ (except $P_1=P_2=0$). Instead, we find steady states for the fractional probabilities $P_i/\PTOT$ (see SI: Processivity), for total remaining probability $\PTOT=\sum_i P_i$.  This determines the dynamics of $\PTOT$, and thus the average fluxes in the cycle at all times $t$: $\PTOT(t) = \exp[-k_{\text{esc}}(P_2/\PTOT)t]$.

Unlike a cycle without escape, in the steady state of $P_i/\PTOT$ the fluxes for the different transitions will not generally be equal. For the two-state cycle, the changes in probabilities $P_1, P_2$ are determined by the flux into and out of the respective states: 
\begin{subequations}
\label{eq:probFlux}
\begin{align}
\frac{\MD P_1}{\MD t} &= J_{21} - J_{12} \\
\frac{\MD P_2}{\MD t} &= J_{12} - J_{21} - J_{\text{esc}} \ .
\end{align}
\end{subequations}
Once $P_i/\PTOT$ reaches steady state, $\MD P_i/\MD t = P_i \ \MD \ln \PTOT / \MD t$. Substituting this into Eqs.~\ref{eq:probFlux} and rearranging gives the pathway (or one-sided) steady-state flux
\begin{equation}
J_{12} = J_{21} + \frac{P_1}{P_1 + P_2}J_{\text{esc}} \ .
\end{equation}

\subsection{Accumulated flux}

Because the flux decays with time as probability escapes, we additionally include the rate of escape in our evaluation of molecular machine progress. Instead of flux alone, we combine flux and (avoided) escape into the `accumulated flux', 
\begin{equation}
\label{eq:TotalFlow}
\Phi_{\text{acc}}(t) = \int_0^t[J_{12}(t') + J_{21}(t')] \ \MD t' \ .
\end{equation}
For $t\to\infty$, $\Phi_{\text{acc}}$ can always be increased by adjusting $\DISS_{ij}$ to reduce $P_2/\PTOT$ and thus reduce escape, so $\Phi_{\text{acc}}(\infty)$ has no maximum, hence we maximize $\Phi_{\text{acc}}(t)$ after a finite time $t$.

\begin{figure}[t!] 
 	\centering
 	\hspace{-0.0in}
 	\begin{tabular}{c}
         \hspace{-0.200in}\includegraphics[width=3.375in]{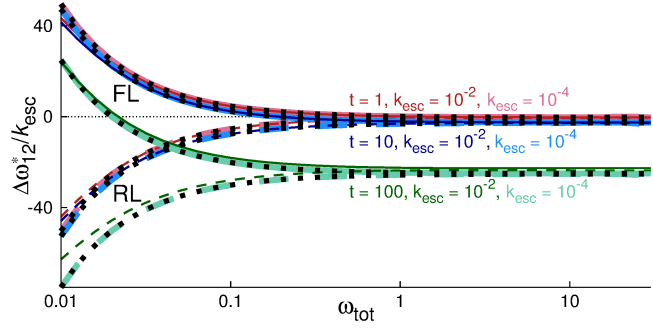}\\
 	\end{tabular}
 	\caption{\label{fig:Escape}  
{\bf Allocating free energy to maximize processivity.}
Free energy allocation $\DISS_{12}$ that maximizes accumulated flux $\Phi_{\text{acc}}$ (Eq.~\ref{eq:TotalFlow}) over varying integration times $t$ (indicated by curve color), in a FL (solid lines) or RL (dashed) two-state cycle with $\BAREONETWO = \BARETWOONE = 1$ and escape rate $k_{\text{esc}}=10^{-2}$ or $10^{-4}$.
Thick black dotted curves show Eqs.~\ref{eq:escape_fl} and \ref{eq:escape_rl}, which approximate $\DELD^*_{12}/k_{\text{esc}}$ with low $k_{\text{esc}}$ for FL and RL, respectively.
Allocation is shown as the difference $\DELD^*_{ij} = \DOPT_{ij} - \frac{1}{2}\TOT$ from the `naive' allocation of equal free energy $\frac{1}{2}\TOT$ to each transition. $\Delta\DISS_{21}^* = -\Delta\DISS_{12}^*$ is not shown.
Initial condition is steady state for $P_i/\PTOT$.
}
\end{figure}

For simplicity, we consider two-state cycles with a single splitting factor $\delta_{ij}^k = \delta$. We set bare rate constants $\BAREONETWO = \BARETWOONE = 1$, such that all escape rate constants $k_{\text{esc}}$ are in units of this bare rate.
Fig.~\ref{fig:Escape} shows two-state free energy allocations that maximize accumulated flux $\Phi_{\text{acc}}(t)$ for FL and RL schemes at various finite times $t$, for two distinct escape rate constants $k_{\text{esc}}$ (see SI: Processivity for similar three-state results).
These escape rate constants are consistent with modeling that suggests the detachment timescale from the vulnerable state of myosin V is 10-1000$\times$ slower than other timescales in the main forward pathway at zero force~\cite{hinczewski13}.

A prominent feature of Fig.~\ref{fig:Escape} is the approximate collapse of cycles with different $k_{\text{esc}}$ to the same $\Delta\DOPT_{12}/k_{\text{esc}}$ (where $\Delta\DOPT_{12} = \DISS_{12} - \tfrac{1}{2}\TOT$), demonstrating that the optimal free energy allocation $\Delta\DOPT_{12} \propto k_{\text{esc}}$ for the parameter values shown in Fig.~\ref{fig:Escape} (SI: Processivity shows optimal allocations for larger $t$, which do not collapse). For FL cycles with small $k_{\text{esc}}$ and $t$,
\begin{equation}
\label{eq:escape_fl}
\frac{\Delta\DOPT_{12}}{k_{\text{esc}}} \simeq -\frac{1}{4}\left(t - \frac{1}{e^{\frac{1}{2}\TOT} - 1}\right) \ ,
\end{equation}
and for RL,
\begin{equation}
\label{eq:escape_rl}
\frac{\Delta\DOPT_{12}}{k_{\text{esc}}} \simeq -\frac{1}{4}\left(t + \frac{1}{1 - e^{-\frac{1}{2}\TOT}}\right) \ .
\end{equation}
See SI: Processivity for derivations. In Fig.~\ref{fig:Escape}, Eqs.~\ref{eq:escape_fl} and \ref{eq:escape_rl} (thick black dotted curves) closely match the numerical $\Delta\DOPT_{12}/k_{\text{esc}}$ (solid and dashed curves) for low $k_{\text{esc}}$ and low $t$.

For high free energy budget $\TOT$, the allocation $\DOPT_{12}$ in Fig.~\ref{fig:Escape} is intuitive: $\DOPT_{12}$ is low and (although not shown in the figure) $\DOPT_{21}=\TOT-\DOPT_{12}$ is high, reducing the probability of being in the vulnerable state 2. For higher escape rates and longer times, the difference $|\DELD^*_{ij}|$ between optimal and naive allocations increases, diverging as time $t\to\infty$.

In Fig.~\ref{fig:Escape}, $\DELD^*_{12}$ for FL changes sign from negative to positive as $\TOT$ decreases, with the change occurring at larger $\TOT$ for shorter times. Positive $\DELD^*_{12}$ increases the probability in the vulnerable state and increases the escape rate.
But maximizing $\Phi_{\text{acc}}$ is a tradeoff between reducing escape and increasing ongoing flux. For longer times $t$ and higher $\TOT$, it is more important to reduce escape to preserve the future flux, and the intuitive result ($\DELD^*_{12} < 0$) is observed. However, at small $t$ and $\TOT$, initial flux is relatively more important, as little escape occurs before time $t$ is reached. The escaping flux from state 2 causes $P_1 > P_2$ for naive free energy allocation $\DISS_{12} = \DISS_{21} = \frac{1}{2}\TOT$. FL only allows the forward fluxes to be increased --- to maximize the flux, the forward flux for transition 12 should be increased more, because $P_1>P_2$, leading to $\DELD^*_{12} > 0$. In contrast, RL only allows reverse fluxes to be decreased --- to maximize the flux, the reverse flux for transition 21 should be decreased more, again because $P_1 > P_2$, consistently leading to $\DELD^*_{12} < 0$.

\section{Discussion}

Biomolecular machines perform a variety of tasks inside cells, driven by the free energy stored in intracellular chemical concentrations. Here we investigated how to maximize flux in molecular machines with multiple \FECS, and how to maintain processivity.

Central to our approach is the general manner in which forward and reverse rate constants are affected by free energy changes~\cite{brown17,schmiedl08,wagoner16}.
\FECSCAP that only affect the forward or reverse rate constants represent extremes, and the division of the effect of \FECS between the forward and reverse rate constants is more generally described by a splitting factor $\delta\in[0,1]$. We primarily consider the simplifying case where splitting factors for all distinct \FECS are equal.

In contrast with previous results~\cite{wagoner16}, we find no single splitting factor $\delta$ generally maximizes the flux. Instead, the optimal splitting factor depends on the detailed free energy allocation, with flux decreasing significantly away from the optimal splitting factor (Fig.~\ref{fig:TwoStateSensitivityDelta}).

Previous studies~\cite{wagoner16,schmiedl08} argued that low splitting factors are optimal, suggesting low splitting factors would describe the operation of evolved biomolecular machines. We have shown that both low and high splitting factors can maximize flux, and that the optimal splitting factor value depends on the allocation of free energy.
Experimental fits of splitting factors find both low splitting factors~\cite{zimmermann12, fisher01, kolomeisky03, hwang17, schmiedl08} and high splitting factors~\cite{fisher01, liepelt07}. 
These fits use several models for the dependence of rate constants on resisting forces~\cite{zimmermann12, fisher01, lau07, kolomeisky03, hwang17, schmiedl08, liepelt07}.
Unlike in our model, all other models we have found assume that rate constants only explicitly contain splitting factors for force, with no splitting factors for other \FECS (though one other model has implicit splitting of the effect of chemical potential~\cite{schmiedl08}). Splitting factor analysis has largely been done in the context of `canonical' molecular machines, such as the walking motors kinesin~\cite{fisher01, lau07, hwang17, liepelt07}, myosin~\cite{kolomeisky03, schmiedl08}, and dynein~\cite{singh05}, or the rotary motor F1-ATPase~\cite{zimmermann12}. Splitting factors $\delta$ are not always robust, with distinct data sets for kinesin motility~\cite{visscher99,carter05} leading to substantially different inferred values of 0.3 and 0.65~\cite{liepelt07}.  It would be an interesting follow up to investigate splitting factors for all \FECS affecting rate constants, and for a broader class of molecular machines, which may lead to a broader diversity of splitting factor values.
In SI: Experimental splitting factors, we describe in more detail the fitted splitting factors for various models~\cite{zimmermann12, fisher01, lau07, kolomeisky03, hwang17, schmiedl08, liepelt07}.

Flux can also be maximized by varying \FECS $\DISS_{ij}^k$, with flux significantly decreasing away from the optimal $\DISS_{ij}^{k*}$. For a scenario with two distinct \FECS, one variable and one fixed, the flux is maximized by allocating the variable \FEC to compensate for any departure from optimal of the fixed \FEC allocation.
There are examples of machine models fit to experimental data where transitions with a fixed load appear to receive more of the other \FECS than those without a load (consistent with our predictions)~\cite{fisher01,clancy11,hwang17}, but also counter-examples that do not fit this optimal framework~\cite{lau07,kolomeisky03}.

Through mutation, biomolecular machine operation can change, and we expect evolution to select machine parameters that favor certain performance characteristics. We argue that robustness increases when tuning parameters to maximize flux with a load, rather than without a load, because for load-optimized parameters the flux is less sensitive to the presence or absence of load. This is consistent with observations of kinesin maintaining a stable velocity under a range of forces~\cite{wang17}, an intuitively beneficial trait~\cite{mukherjee17}.

We also examined cycles that are transiently processive and escape from a single vulnerable state. For large free energy budgets, maximizing the number of complete cycles before the cycle escapes leads to intuitive allocations: free energy is primarily allocated to decrease the probability present in the vulnerable state. However, for small free energy budgets, to maximize the number of cycles completed, free energy is primarily allocated to increase the flux --- for forward labile cycles this increases the probability present in the vulnerable state, while for reverse labile cycles this decreases it. Although this forward labile allocation increases the rate of escape, the associated flux increase more than compensates, leading to an overall increase in the accumulated flux.

Kinesins mutated to have different neck linker charge show increased processivity, which is attributed to shorter waiting times in states vulnerable to detachment from the microtubule~\cite{andreasson15,thorn00,toprak09}. These findings are consistent with our result that occupation of the vulnerable state be reduced to maximize processivity for large free energy budgets, such as the $\sim 20k_{\text{B}}T$ of free energy from ATP hydrolysis~\cite{phillips12} driving each kinesin cycle~\cite{schnitzer97}.

While we are unaware of other modeling that maximizes processivity, 
Hill modeled escape as a transition to the initial state where all trajectories begin, effectively acting as an additional subcycle~\cite{hill88}. This steady state with immediate rebinding is distinct from our steady state without rebinding.  Both Hill's approach and ours allow calculation of detachment rates (distinct due to differing steady states); but our approach additionally quantifies progress before detachment.

\begin{acknowledgement}

This work was supported by a Natural Sciences and Engineering Research Council of Canada (NSERC) Discovery Grant, by funds provided by the Faculty of Science, Simon Fraser University through the President's Research Start-up Grant, by a Tier II Canada Research Chair, and by WestGrid (www.westgrid.ca) and Compute Canada Calcul Canada (www.computecanada.ca). 
The authors thank Emma Lathouwers and Steven Large (SFU Physics) for useful discussions and feedback.

\end{acknowledgement}

\section{Supplemental Information}

\renewcommand\thefigure{S\arabic{figure}}
\renewcommand\thetable{S\arabic{table}}
\renewcommand\theequation{S\arabic{equation}}
\setcounter{equation}{0}
\setcounter{figure}{0}
\setcounter{table}{0}

\subsection{Maximizing flux: Varying splitting factors}
\label{sec:load_app}

\FECSCAP $\DISS_{ij}^k$ influence the forward and reverse rate constants, parameterized by $\delta_{ij}^k$,
\begin{equation}
\label{eq:RateConstantsOne}
\RCF = \BARE e^{\sum_k \delta_{ij}^k\sigma_{ij}^k} \ \ \text{and} \ \  \RCR = \BARE e^{-\sum_k(1 - \delta_{ij}^k)\DISS_{ij}^k} \ .
\end{equation}
In the main text we showed that if $\delta_{ij}^k = \delta$, there is no value of $\delta$ that generally maximizes the flux. The results of this section, with no consistent sign for $\partial J/\partial\SPLITONE_{12}$, $\partial J/\partial\SPLITONE$, and $\partial J/\partial\delta_{12}$, similarly conclude that there are no generally optimal values for $\delta_{ij}^k$. Having no general optimal $\delta_{ij}^k$ value does not align with previous results~\cite{wagoner16} where the power of a motor was maximized for a power stroke ($\delta = 0$).

The two-state cycle steady-state flux is
\begin{equation}
J = \frac{\RCFONETWO\RCFTWOONE - \RCRONETWO\RCRTWOONE}{\RCFONETWO + \RCRONETWO + \RCFTWOONE + \RCRTWOONE} \ .
\end{equation}
For the cases below we consider two \FECS, $\DISSONE$ and $\DISSTWO$, subject to the relationships $\TOTONE = \sum\DISSONE_{ij}$, $\TOTTWO = \sum\DISSTWO_{ij}$, and $\TOT = \TOTONE + \TOTTWO$.

\subsubsection{Case 1}

In this case, 
we allow the splitting constants to differ between transitions, but require equality for different \FECS, $\delta_{ij}^k = \delta_{ij}$. This gives flux
\begin{equation}
J = \dfrac{\BAREONETWO\BARETWOONE(1 - e^{-\TOT})}{\BAREONETWO e^{-\delta_{21}\DISS_{21}}(1 + e^{-\DISS_{12}}) + \BARETWOONE e^{-\delta_{12}\DISS_{12}}(1 + e^{-\DISS_{21}})} \ .
\end{equation}
The flux increases with increasing $\delta_{12}$ as
\begin{align}
\frac{\partial J}{\partial \delta_{12}} = \dfrac{\BAREONETWO(\BARETWOONE)^2\DISS_{12} (1 - e^{-\TOT}) e^{-\delta_{12}\DISS_{12}}(1 + e^{-\DISS_{21}}) }{\left[\BAREONETWO e^{-\delta_{21}\DISS_{21}}(1 + e^{-\DISS_{12}}) + \BARETWOONE e^{-\delta_{12}\DISS_{12}}(1 + e^{-\DISS_{21}})\right]^2} \ .
\end{align}
The flux is maximized for $\delta_{12} = 0$ if $\DISS_{12} < 0$ and for $\delta_{12} = 1$ if $\DISS_{12} > 0$.  Variation of $\delta_{21}$ produces a similar result.

\subsubsection{Case 2}

If we allow the splitting constants to differ for different \FECS, but not between transitions, so that $\SPLITONE_{ij} = \SPLITONE$ and $\SPLITTWO_{ij} = \SPLITTWO$ in Eq.~\ref{eq:RateConstantsOne}, the flux and its derivative are
\begin{subequations}
\begin{align}
\label{eq:AppFlux1}
J &= \dfrac{\BAREONETWO\BARETWOONE(1 - e^{-\TOT})}{\BAREONETWO e^{-\SPLITONE\DISSONE_{21} - \SPLITTWO \DISSTWO_{21}}(1 + e^{-\DISS_{12}}) + \BARETWOONE e^{-\SPLITONE\DISSONE_{12} - \SPLITTWO \DISSTWO_{12}}(1 + e^{-\DISS_{21}})} \ . \\
\frac{\partial J}{\partial \SPLITONE} &= \dfrac{\BAREONETWO\BARETWOONE(1 - e^{-\TOT})\left[\BAREONETWO\DISSONE_{21} e^{-\SPLITONE\DISSONE_{21} - \SPLITTWO\DISSTWO_{21}}(1 + e^{-\DISS_{12}}) + \BARETWOONE\DISSONE_{12} e^{-\SPLITONE\DISSONE_{12} - \SPLITTWO \DISSTWO_{12}}(1 + e^{-\DISS_{21}}) \right]}{\left[\BAREONETWO e^{-\SPLITONE\DISSONE_{21} - \SPLITTWO \DISSTWO_{21}}(1 + e^{-\DISS_{12}}) + \BARETWOONE e^{-\SPLITONE\DISSONE_{12} - \SPLITTWO \DISSTWO_{12}}(1 + e^{-\DISS_{21}})\right]^2} \ .
\end{align}
\end{subequations}
If $\DISSONE_{12}>0$ and $\DISSONE_{21} > 0$ then $\partial J/\partial\SPLITONE > 0$, and $\SPLITONE=1$ maximizes the flux.
However, both $\DISSONE_{12}$ and $\DISSONE_{21}$ are not always positive, so $\partial J/\partial \SPLITONE$ is not always positive or negative, with the flux maximized for
\begin{align}
\SPLITONE =& \frac{1}{\DISSONE_{21} - \DISSONE_{12}} \left\{\SPLITTWO(\DISSTWO_{12} - \DISSTWO_{21}) +\ln\left[\frac{-\DISSONE_{21}\BAREONETWO(1 + e^{-\DISS_{12}})}{\DISSONE_{12}\BARETWOONE(1 + e^{-\DISS_{21}})}\right]\right\} \ .
\end{align}
Similar results are found for variation of $\SPLITTWO$.

\subsubsection{Case 3}

The most general case has splitting constants varying between transitions and between \FECS, as in Eq.~\ref{eq:RateConstantsOne}, giving
\begin{subequations}
\begin{align}
J &= \dfrac{\BAREONETWO\BARETWOONE(1 - e^{-\TOT})}{\BAREONETWO e^{-\SPLITONE_{21}\DISSONE_{21} - \SPLITTWO_{21} \DISSTWO_{21}}(1 + e^{-\DISS_{12}}) + \BARETWOONE e^{-\SPLITONE_{12}\DISSONE_{12} - \SPLITTWO_{12} \DISSTWO_{12}}(1 + e^{-\DISS_{21}})} \ , \\
\frac{\partial J}{\partial \SPLITONE_{12}} &= \dfrac{\BAREONETWO(\BARETWOONE)^2\DISSONE_{12}(1 - e^{-\TOT}) e^{-\SPLITONE_{12}\DISSONE_{12} - \SPLITTWO_{12}\DISSTWO_{12}}(1 + e^{-\DISS_{12}})}{\left[\BAREONETWO e^{-\SPLITONE_{21}\DISSONE_{21} - \SPLITTWO_{21} \DISSTWO_{21}}(1 + e^{-\DISS_{12}}) + \BARETWOONE e^{-\SPLITONE_{12}\DISSONE_{12} - \SPLITTWO_{12} \DISSTWO_{12}}(1 + e^{-\DISS_{21}})\right]^2} \ .
\end{align}
\end{subequations}

If $\DISSONE_{12} > 0$, then $\partial J/\partial\SPLITONE_{12} > 0$, leading to $\SPLITONE_{12} = 1$ maximizing flux; conversely $\DISSONE_{12} < 0$ leads to $\SPLITONE_{12} = 0$ maximizing flux. Similar results are found for variation of $\SPLITONE_{21}$, $\SPLITTWO_{12}$, and $\SPLITTWO_{21}$.

\subsection{Maximizing flux: Varying free energy}

Here we examine what free energy allocation $\DISS_{ij}$ (composed of \FECS $\DISS_{ij}^k$) maximizes the flux. We go through distinct cases for splitting the effect of two \FECS between forward and reverse rate constants. In the main text, we find that for free energy components split identically across all transitions ($\SPLITONE_{ij} = \SPLITTWO_{ij} = \delta$), there is no closed form for the optimal free energy allocation $\DOPT_{ij}$, but that for $\delta = 0$ or $\delta = 1$, closed forms can be found. We review these results below, as well as other splitting scenarios with similar results.

\subsubsection{Case 1}

Here both \FECS are split identically across all transitions, $\SPLITONE_{ij} = \SPLITTWO_{ij} = \delta$, 
giving rate constants
\begin{equation}
\RCF = \BARE e^{\delta \DISS_{ij}} \ \ \text{and} \ \  \RCR = \BARE e^{-(1-\delta)\DISS_{ij}} \ .
\end{equation}
The flux and its derivative are
\begin{subequations}
\begin{align}
J &= \dfrac{\BAREONETWO\BARETWOONE e^{\delta \TOT}(1 - e^{-\TOT})}{\BAREONETWO e^{\delta \DISS_{12}}(1 + e^{-\DISS_{12}}) + \BARETWOONE e^{\delta(\TOT - \DISS_{12})}(1 + e^{-(\TOT - \DISS_{12})})} \ , \\
\frac{\partial J}{\partial \DISS_{12}} &= \dfrac{-\BAREONETWO\BARETWOONE e^{\delta \TOT}(1 - e^{-\TOT}) \left\{\BAREONETWO e^{\delta \DISS_{12}}\left[\delta - (1-\delta)e^{-\DISS_{12}}\right] + \BARETWOONE e^{\delta(\TOT - \DISS_{12})}\left[-\delta + (1-\delta)e^{-(\TOT - \DISS_{12})}\right]\right\}}{\left[\BAREONETWO e^{\delta \DISS_{12}}(1 + e^{-\DISS_{12}}) + \BARETWOONE e^{\delta(\TOT - \DISS_{12})}(1 + e^{-(\TOT - \DISS_{12})})\right]^2} \ .
\end{align}
\end{subequations}

Setting $\partial J/\partial \DISS_{12} = 0$ yields
\begin{align}
\label{eq:ConditionOne}
\BAREONETWO e^{\delta \DISS_{12}^*}\left[\delta - (1-\delta)e^{-\DISS_{12}^*}\right] = \BARETWOONE e^{\delta(\TOT - \DISS_{12}^*)}\left[\delta - (1-\delta)e^{-(\TOT - \DISS_{12}^*)}\right] \ .
\end{align}

\subsubsection{Case 2}

For this case two \FECS are split differently, but the splitting is the same for all transitions. The rate constants are then
\begin{equation}
\RCF = \BARE e^{\SPLITONE\DISSONE_{ij} + \SPLITTWO \DISSTWO_{ij}} \ \ \text{and} \ \  \RCR = \BARE e^{-(1-\SPLITONE)\DISSONE_{ij} - (1 - \SPLITTWO)\DISSTWO_{ij}} \ .
\end{equation}
The flux and its derivative are
\begin{subequations}
\begin{align}
J &= \dfrac{\BAREONETWO\BARETWOONE e^{\SPLITONE\TOTONE + \SPLITTWO \TOTTWO}(1 - e^{-\TOT})}{\left\{\BAREONETWO e^{\SPLITONE \DISSONE_{12} + \SPLITTWO\DISSTWO_{12}}(1 + e^{-\DISS_{12}}) + \BARETWOONE e^{\SPLITONE(\TOTONE - \DISSONE_{12}) + \SPLITTWO (\TOTTWO - \DISSTWO_{12})}\left(1 + e^{-(\TOT - \DISS_{12})}  \right)\right\}} \ , \\
\frac{\partial J}{\partial \DISSONE_{12}} &= \dfrac{-\BAREONETWO\BARETWOONE e^{\SPLITONE\TOTONE + \SPLITTWO \TOTTWO}(1 - e^{-\TOT})\left\{\splitdfrac{\BAREONETWO e^{\SPLITONE\DISSONE_{12} + \SPLITTWO \DISSTWO_{12}}\left[\SPLITONE - (\SPLITONE - 1)e^{-\DISS_{12}}\right] + }{\BARETWOONE e^{\SPLITONE(\TOTONE-\DISSONE_{12}) + \SPLITTWO(\TOTTWO - \DISSTWO_{12})}\left[-\SPLITONE + (1-\SPLITONE)e^{-(\TOT - \DISS_{12})}\right]}\right\}}{\left\{\BAREONETWO e^{\SPLITONE \DISSONE_{12} + \SPLITTWO\DISSTWO_{12}}(1 + e^{-\DISS_{12}}) + \BARETWOONE e^{\SPLITONE(\TOTONE - \DISSONE_{12}) + \SPLITTWO (\TOTTWO - \DISSTWO_{12})}\left(1 + e^{-(\TOT - \DISS_{12})}  \right)\right\}^2} \ .
\end{align}
\end{subequations}

\subsubsection{Case 3}

For this case, different \FECS are split the same, but the splitting is different for each transition. The rate constants are then
\begin{equation}
\RCF = \BARE e^{\delta_{ij}\DISS_{ij}} \ \ \text{and} \ \  \RCR = \BARE e^{-(1-\delta_{ij})\DISS_{ij}} \ .
\end{equation}
The flux and its derivative are
\begin{subequations}
\begin{align}
J &= \dfrac{\BAREONETWO\BARETWOONE (1 - e^{-\TOT})}{\left[\BAREONETWO e^{-\delta_{21} (\TOT - \DISS_{12})}(1 + e^{-\DISS_{12}}) + \BARETWOONE e^{-\delta_{12}\DISS_{12}}\left(1 + e^{-(\TOT-\DISS_{12})}  \right)\right]} \ , \\
\frac{\partial J}{\partial \DISS_{12}} &= \dfrac{-\BAREONETWO\BARETWOONE (1 - e^{-\TOT})\left\{\splitdfrac{\BAREONETWO e^{-\delta_{21} (\TOT-\DISS_{12})}\left[\delta_{21} + (\delta_{21} - 1)e^{-\DISS_{12}}\right] - }{\BARETWOONE e^{-\delta_{12}\DISS_{12}} \left[\delta_{12} + (\delta_{12} - 1)e^{-(\TOT-\DISS_{12})}\right]}\right\}}{\left[\BAREONETWO e^{-\delta_{21} (\TOT - \DISS_{12})}(1 + e^{-\DISS_{12}}) + \BARETWOONE e^{-\delta_{12}\DISS_{12}}\left(1 + e^{-(\TOT-\DISS_{12})}  \right)\right]^2}  \ .
\end{align}
\end{subequations}

\subsubsection{Case 4}

For this case, different \FECS are split differently, and the splitting is different for each transition. The rate constants are then
\begin{equation}
\RCF = \BARE e^{\SPLITONE_{ij}\DISSONE_{ij} + \SPLITTWO_{ij} \DISSTWO_{ij}} \ \ \text{and} \ \  \RCR = \BARE e^{-(1-\SPLITONE_{ij})\DISSONE_{ij} - (1 - \SPLITTWO_{ij})\DISSTWO_{ij}} \ .
\end{equation}
The flux and its derivative are
\begin{subequations}
\begin{align}
J &= \dfrac{\BAREONETWO\BARETWOONE (1 - e^{-\TOT})}{\BAREONETWO e^{-\SPLITONE_{21}(\TOTONE - \DISSONE_{12}) - \SPLITTWO_{21}(\TOTTWO - \DISSTWO_{12})}(1 + e^{-\DISS_{12}}) + \BARETWOONE e^{-\SPLITONE\DISSONE_{12} - \SPLITTWO\DISSTWO_{12}}\left(1 + e^{-(\TOT - \DISS_{12})}\right)} \ , \\
\label{eq:casefour}
\frac{\partial J}{\partial \DISSONE_{12}} &= \dfrac{-\BAREONETWO\BARETWOONE (1 - e^{-\TOTONE - \TOTTWO})\left\{\splitdfrac{\BAREONETWO e^{-\SPLITONE_{21}(\TOTONE-\DISSONE_{12}) - \SPLITTWO_{21}(\TOTTWO - \DISSTWO_{12})}\left[\SPLITONE_{21} + (\SPLITONE_{21} - 1)e^{-\DISS_{12}}\right] - }{\BARETWOONE e^{-\SPLITONE_{12}\DISSONE_{12} - \SPLITTWO_{12}\DISSTWO_{12}}\left[\SPLITONE_{12} + (\SPLITONE_{12} - 1)e^{-(\TOT - \DISS_{12})}\right]}\right\}}{\left[\BAREONETWO e^{-\SPLITONE_{21}(\TOTONE - \DISSONE_{12}) - \SPLITTWO_{21}(\TOTTWO - \DISSTWO_{12})}(1 + e^{-\DISS_{12}}) + \BARETWOONE e^{-\SPLITONE\DISSONE_{12} - \SPLITTWO\DISSTWO_{12}}\left(1 + e^{-(\TOT - \DISS_{12})}\right)\right]^2} \ .
\end{align}
\end{subequations}

Case 1 is considered in the main text, while cases 2-4 are not. A closed form for $\DISS_{12}^*$ (when splitting factors are equal for different \FECS) or $\DISSONEOPT_{12}$ (when they are not)
solving $\partial J/\partial \DISS_{12} = 0$ or $\partial J/\partial \DISSONE_{12} = 0$, respectively, cannot generally be found for any of these cases.
We now consider some simplifying cases, with one or both of splitting factors $\SPLITONE$ and $\SPLITTWO$ set to the extremes 0 or 1, to arrive at some closed-form solutions for $\DISS_{12}^*$.

\subsubsection{Case 5}
For this case, all \FECS split the same, and the splitting is the same for each transition, with splitting factor $\delta = 1$, corresponding to forward labile. 
Setting $\delta = 1$ in Eq.~\ref{eq:ConditionOne} gives the flux-maximizing free energy allocation
\begin{equation}
\DISS_{12}^* = \frac{1}{2}\TOT + \frac{1}{2}\ln\frac{\BARETWOONE}{\BAREONETWO} \ .
\end{equation}
This corresponds to
\begin{equation}
\DISSONEOPT_{12} = -\DISSTWOOPT_{12} + \frac{1}{2}\TOT + \frac{1}{2}\ln\frac{\BARETWOONE}{\BAREONETWO} \ .
\end{equation}

\subsubsection{Case 6}
For this case, all \FECS split the same, and the splitting is the same for each transition, with splitting factor $\delta = 0$ corresponding to reverse labile. 
Setting $\delta = 0$ in Eq.~\ref{eq:ConditionOne} gives the flux-maximizing free energy allocation
\begin{equation}
\DISS_{12}^* = \frac{1}{2}\TOT - \frac{1}{2}\ln\frac{\BARETWOONE}{\BAREONETWO} \ .
\end{equation}
This corresponds to
\begin{equation}
\DISSONEOPT_{12} = -\DISSTWOOPT_{12} + \frac{1}{2}\TOT - \frac{1}{2}\ln\frac{\BARETWOONE}{\BAREONETWO} \ .
\end{equation}

For cases 5 and 6 the flux-maximizing allocation of $\DISSONE$ compensates exactly for $\DISSTWO$ on each transition, and uses the remaining free energy to maximize the flux as if $\DISSTWO=0$.

\subsection{Experimental splitting factors}

Table~\ref{tab:splittingfactors} summarizes splitting factors derived from fits to experimental data.

\begin{table}[tbp]
  \centering
  \begin{tabular}{cccc}
  $\delta$ range & Model & Machine\\
  \hline
  \hline
  \rule{0pt}{3ex} $0.004-0.024$ & Eq.~\ref{eq:schmiedl} & Myosin\cite{schmiedl08}\\
  $-0.03 - 0.08$ & Eq.~\ref{eq:fisher} & Myosin\cite{kolomeisky03}\\
  $0 - 0.13$ & Eq.~\ref{eq:fisher} & Kinesin\cite{hwang17}\\
  $0.1-0.3$ & Eq.~\ref{eq:zimmermann} & F$_1$ ATPase\cite{zimmermann12}\\
  $0.04-0.48$ & Eq.~\ref{eq:fisher} & Kinesin (four-state)\cite{fisher01} \\
  $0.04-0.62$ & Eq.~\ref{eq:fisher} & Kinesin (two-state)\cite{fisher01} \\
  $0.3-0.65$ & Eq.~\ref{eq:liepelt} & Kinesin\cite{liepelt07}\\
  $0.04-2.8$ & Eq.~\ref{eq:fisher} & Kinesin\cite{lau07}\\
  \end{tabular}
  \caption{\label{tab:splittingfactors} 
  {\bf Splitting factors from experimental fits}.
  Splitting factors $\delta$ fit to data from a variety of molecular machines, arranged in order of range midpoints.}
\end{table}

In Schmiedl and Seifert~\cite{schmiedl08} a discrete-state model has rate constants
\begin{subequations}
\label{eq:schmiedl}
\begin{align}
k^+ &= k^0 e^{\Delta\mu^+ - \beta\delta Fd} \ , \\
k^- &= k^0 e^{-\Delta\mu^- + \beta(1-\delta) Fd} \ .
\end{align}
\end{subequations}
Their $\delta$ are identical to our splitting factors. For a two-state myosin model, they find $\delta_{12} = 0.004$ and $\delta_{21} = 0.024$.

Zimmermann and Seifert~\cite{zimmermann12} model a continuous free energy landscape with forward and reverse rate constants
\begin{subequations}
\label{eq:zimmermann}
\begin{align}
k^+(n,x) &= k^{0}e^{\Delta\mu_{\text{ATP}} - \kappa d\delta (d\delta/2 + n - x)} \ , \\
k^-(n,x) &= k^{0}e^{\Delta\mu_{\text{ADP}} + \Delta\mu_{\text{P$_i$}} - \kappa d(1-\delta)[d(1-\delta)/2 - n + x]} \ .
\end{align}
\end{subequations}
$d$ is motor step size, $\kappa$ is the spring constant between the motor and attached probe, and $\Delta\mu_i = \mu_i - \mu_i^{\text{eq}} = \ln c_i/c_i^{\text{eq}}$ are the chemical potential differences from equilibrium. Rate constants $k^{+/-}$ and $k^0$ do not have indices because the model contains only one step. Splitting factor $\delta$ is found by fitting to experimental F$_1$ ATPase data.

In Liepelt and Lipowsky~\cite{liepelt07} a discrete-state model has rate constants
\begin{equation}
\label{eq:liepelt}
k_{ij} = k_{ij}^0\Phi(F) \ .
\end{equation}
For chemical transitions, $\Phi_{ij}(F) = \Phi_{ji}(F) = 2/[1 + e^{\chi_{ij} F/k_{\text{B}}T}]$, with $\chi_{ij}$ a dimensionless force parameter similar to $\delta$ or $\theta$. For mechanical transitions, $\Phi(F) = e^{-\delta F}$ for the direction against the load, and $\Phi(F) = e^{(1 - \delta)F}$ for the opposite direction. Using distinct kinesin motility data sets~\cite{visscher99,carter05}, they find $\delta = 0.3$ or 0.6.

Fisher and Kolomeisky~\cite{fisher01} construct a discrete-state model with forward and reverse rate constants
\begin{subequations}
\label{eq:fisher}
\begin{align}
k_{ij}^+ &= k_{ij}^{+0} e^{-\theta_{ij}^+ Fd} \ , \\
k_{ij}^- &= k_{ij}^{-0} e^{\theta_{ij}^- Fd} \ .
\end{align}
\end{subequations}
$F$ is a constant opposing force, $d$ is the motor step over a complete cycle, and $k_{ij}^{\pm 0}$ are the rate constants at zero force. The $\theta_{ij}^{+/-}$ are found by fitting to experimental kinesin data, but are not identical to our splitting factors $\delta_{ij}$. Instead, they use the constraint $\sum(\theta_{ij}^+ + \theta_{ij}^-) = 1$, such that $\theta_{ij}^{+/-}$ is a combination of allocating a load to the various transitions and splitting the impact of the load between the forward and reverse rate constants. 
For a two-state model, for the first transition, they find $\theta_{12}^+ = 0.135$ and $\theta_{12}^- = 0.08$, equivalent to $\delta_{12} = 0.62$. For the second transition, they find $\theta_{21}^+ = 0.035$ and $\theta_{21}^- = 0.75$, equivalent to $\delta_{21} = 0.04$. For a four-state model, the first transition is found to have $\theta_{12}^+ = 0.12$ and $\theta_{12}^- = 0.13$ ($\delta_{12} = 0.48$), the second transition $\theta_{23}^+ = 0.02$ and $\theta_{23}^- = 0.13$ ($\delta_{23} = 0.13$), the third transition $\theta_{34}^+ = 0.02$ and $\theta_{34}^- = 0.13$ ($\delta_{34} = 0.13$), and the fourth transition $\theta_{41}^+ = 0.02$ and $\theta_{41}^- = 0.43$ ($\delta_{41} = 0.04$). The second transition of the two-state model, and the fourth transition of the four-state model, are treated as the most irreversible, indicating these transitions are allocated the most free energy. These transitions are also fit with the largest $\theta$ values, so that more of the other \FECS are allocated to transitions with more load. A later publication by the same authors~\cite{kolomeisky03}, with a two-state myosin model using the same force dependence, fitted the first transition with $\theta_{12}^+ = -0.01$ and $\theta_{12}^- = 0.385$ ($\delta_{12} = -0.03$) and a second transition with $\theta_{21}^+ = 0.045$ and $\theta_{21}^- = 0.58$ ($\delta_{21} = 0.08$). In this case, the second transition is fit with a greater load, but is not allocated a greater amount of other \FECS.

In Hwang and Hyeon~\cite{hwang17}, a discrete-state model for kinesin has forward and reverse rate constants with the same force dependence as Eq.~\ref{eq:fisher}. With four states, the first transition is fit with $\theta_{12}^+ = 0$ and $\theta_{12}^- = 0.15$ ($\delta_{12} = 0$), the second transition with $\theta_{23}^+ = 0.04$ and $\theta_{23}^- = 0.5$ ($\delta_{23} = 0.07$), the third transition with $\theta_{34}^+ = 0.01$ and $\theta_{34}^- = 0.14$ ($\delta_{34} = 0.07$), and the fourth transition with $\theta_{41}^+ = 0.02$ and $\theta_{41}^- = 0.14$ ($\delta_{41} = 0.125$). The second transition is fit with the largest load (high $\theta$ values) and has a larger amount of the other \FECS. 

In Lau \emph{et al}~\cite{lau07}, a discrete-state model for kinesin has forward and reverse rate constants with the same force dependence as Eq.~\ref{eq:fisher}. With two states, the first transition is fitted with $\theta_{12}^+ = 0.25$ and $\theta_{12}^- = -0.16$ ($\delta_{12} = 2.8$) and the second transition with $\theta_{21}^+ = 0.08$ and $\theta_{21}^- = 1.83$ ($\delta_{21} = 0.04$). These are fit with the constraint $\theta_{12}^+ + \theta_{12}^- + \theta_{21}^+ + \theta_{21}^- = 2$. The transition state picture (\emph{e.g.} in Schmiedl and Seifert~\cite{schmiedl08}) describes an opposing force as reducing the rate of forward transitions and/or increasing the rate of reverse transitions. This transition state picture does not explain the $\theta_{ij}^{+/-}$ values in Lau \emph{et al}~\cite{lau07} because for transition 12 both the forward and reverse transition rates are decreased by opposing force.

\subsection{Processivity}
\label{sec:vulnerable_app}

\subsubsection{Two states}
With state 2 vulnerable to escape, the governing differential equations are
\begin{subequations}
\begin{align}
\label{eq:ShowEscapeA}
\frac{\MD P_1}{\MD t} &= -(\RCFONETWO + \RCRTWOONE)P_1 + (\RCRONETWO + \RCFTWOONE)P_2 \\
\label{eq:ShowEscape}
\frac{\MD P_2}{\MD t} &= (\RCFONETWO + \RCRTWOONE)P_1 - (\RCRONETWO + \RCFTWOONE)P_2 - k_{\text{esc}}P_2 \ .
\end{align}
\end{subequations}
We rewrite these differential equations in terms of $P_i/\PTOT$, where $\PTOT(t) = P_1(t) + P_2(t)$ is the total remaining probability that has not yet escaped the cycle. Using	
\begin{equation}
\label{eq:conversion}
\frac{\MD \left(\frac{P_i}{\PTOT}\right)}{\MD t} = \frac{1}{\PTOT}\frac{\MD P_i}{\MD t} - \frac{P_i}{\PTOT^2} \frac{\MD\PTOT}{\MD t} \ ,
\end{equation}
and $\MD\PTOT/\MD t = -k_{\text{esc}}P_2$, gives
\begin{equation}
\label{eq:escape_de_1}
\frac{\MD\left(\frac{P_1}{\PTOT}\right)}{\MD t} = -(\RCFONETWO + \RCRTWOONE)\frac{P_1}{\PTOT} + (\RCRONETWO + \RCFTWOONE)\frac{P_2}{\PTOT} + k_{\text{esc}}\frac{P_1}{\PTOT}\frac{P_2}{\PTOT} \ .
\end{equation}
Substituting $\tfrac{P_1}{\PTOT} = 1 - \tfrac{P_2}{\PTOT}$ into Eq.~\ref{eq:escape_de_1} at steady state (when $\MD(P_i/\PTOT)/\MD t = 0$) gives a quadratic equation for $P_2/\PTOT$:
\begin{equation}
\label{eq:escape_quadratic}
\begin{split}
k_{\text{esc}}\left(\frac{P_2}{\PTOT}\right)^2 - \left[\RCFONETWO + \RCRONETWO + \RCFTWOONE + \RCRTWOONE + k_{\text{esc}}\right]\left(\frac{P_2}{\PTOT}\right) + \RCFONETWO + \RCRTWOONE = 0 \ .
\end{split}
\end{equation}
$P_2/\PTOT$ has a real solution when the discriminant is non-negative.

\subsubsection{Optimizing two-state flux accumulated over large $t$}

\begin{figure}[t] 
 	\centering
 	\hspace{-0.0in}
 	\begin{tabular}{c}
         \hspace{-0.300in}\includegraphics[width=3.375in]{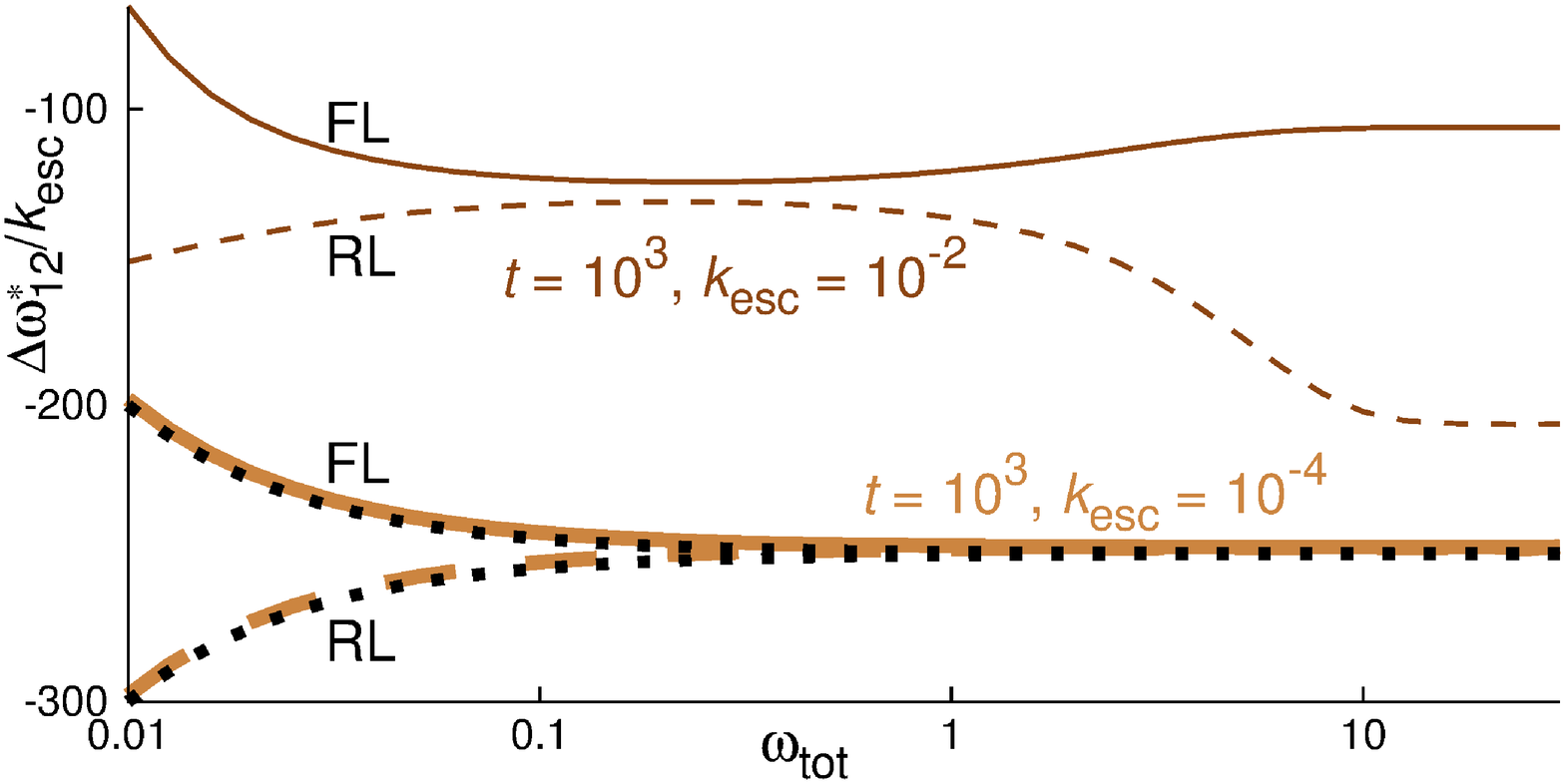}\\
 	\end{tabular}
 	\caption{\label{fig:EscapeAppendix}  
{\bf Non-monotonic optimal free energy allocation reflects competition between flux and avoiding escape.} Free energy allocation $\DISS_{ij}$ that maximizes accumulated flux $\Phi_{\text{acc}}$ (Eq.~16)
for $t=1000$, in two-state cycle with $\BAREONETWO = \BARETWOONE = 1$ and variable escape rate $k_{\text{esc}}$.
FL plotted as solid lines, RL as dashed lines. Thick black dotted curves show Eqs.~17
and 18,
which approximate $\DELD^*_{12}/k_{\text{esc}}$ for low $k_{\text{esc}}$.
Allocation is shown as the difference $\DELD^*_{ij} = \DOPT_{ij} - \tfrac{1}{2}\TOT$ from the `naive' equal allocation $\tfrac{1}{2}\TOT$ to each transition. $\Delta\DISS_{21}^* = -\Delta\DISS_{12}^*$ is not shown. 
Initial condition is the steady state for $P_i/\PTOT$.
}
\end{figure}

Fig.~4
shows that for small $k_{\text{esc}}$ and $t$, $\DELD^*_{12}$ changes monotonically as $\TOT$ increases. However, for $k_{\text{esc}}=10^{-2}$ and $t=1000$, $\DELD^*_{12}$ is non-monotonic (Fig.~\ref{fig:EscapeAppendix}). As mentioned in the main text, the accumulated flux $\Phi_{\text{acc}}$ is the integrated flux over a period of time, and reflects both the instantaneous flux and escape from state 2.
Increasing $\TOT$ shifts the respective functional dependences of flux and $P_2/\PTOT$ on $\DISS_{12}$, often in ways producing competing influences on $\DELD^*_{12}$.
For sufficiently high $\TOT$, the functional dependences asymptote to forms that don't change with further increased $\TOT$. For small $t$ and $k_{\text{esc}}$, these functional dependences cease changing at approximately the same high value of $\TOT$, leading $\DELD^*_{12}$ to be monotonic in $\TOT$ in Fig.~4.
However for larger $t$ and $k_{\text{esc}}$, the functional dependences cease changing at different values of $\TOT$. Since the (potentially countervailing) influences on $\DELD^*_{12}$ do not happen over the same $\TOT$ range, $\DELD^*_{12}$ can change non-monotonically with $\TOT$, as seen for $t=10^3$ and $k_{\text{esc}}=10^{-2}$ in Fig.~\ref{fig:EscapeAppendix}.

For $k_{\text{esc}}=10^{-2}$ and $t=1000$ (Fig.~\ref{fig:EscapeAppendix}), $\DELD^*_{12}$ for FL and RL do not coincide at high $\TOT$. This distinction between FL and RL at high $\TOT$ occurs because with more free energy dissipation, FL increases rate constants while RL decreases rate constants. At high $\TOT$, the rate constants for FL can still significantly change with more $\DISS_{ij}$, while the rate constants for RL are already quite small. This leads RL to require a larger $|\DELD^*_{12}|$ than FL. This effect also occurs for lower $k_{\text{esc}}$ and $t$ in Fig.~4,
but is much smaller because escape rates and/or total probability escaping is much smaller.

\subsubsection{Two-state optimization approximation for low $k_{\text{esc}}$, $t$}

Equations~17
and 18
are approximations for the free energy allocation $\Delta\DOPT_{12}$ that maximize the accumulated flux for small $k_{\text{esc}}$ and $t$.

We first derive Eq.~17
for FL rate constants with $\BAREONETWO = \BARETWOONE = 1$, $\RCFONETWO = e^{\TOT/2}e^{\DELD_{12}}$, and $\RCFTWOONE = e^{\TOT/2}e^{-\DELD_{12}}$,
beginning with Eq.~\ref{eq:escape_quadratic}
and solving for $P_2$ using the quadratic formula,
\begin{equation}
\frac{P_2}{\PTOT} = (e^{\TOT/2}e^{\DELD_{12}} + e^{\TOT/2}e^{-\DELD_{12}} + 2 + k_{\text{esc}}) \left(1 - \sqrt{1 - \frac{4k_{\text{esc}}(e^{\TOT/2}e^{\DELD_{12}} + 1)}{(e^{\TOT/2}e^{\DELD_{12}} + e^{\TOT/2}e^{-\DELD_{12}} + 2 + k_{\text{esc}})^2}}\right) \ .
\end{equation}
Taylor expanding for small $k_{\text{esc}}$ and $\DELD_{12}$ gives
\begin{equation}
\label{eq:p2_approx}
\frac{P_2}{\PTOT} \simeq \frac{1}{2}\left[1 + \frac{\DELD_{12}}{1 + e^{-\TOT/2}} - \frac{k_{\text{esc}}}{4(1 + e^{\TOT/2})}\right] \ .
\end{equation}
The flux is
\begin{equation}
\label{eq:combinedflux}
J_{12} + J_{21} = e^{\TOT/2}e^{\DELD_{12}}\frac{P_1}{\PTOT} - \frac{P_2}{\PTOT} + e^{\TOT/2}e^{-\DELD_{12}}\frac{P_2}{\PTOT} - \frac{P_1}{\PTOT} \ .
\end{equation}
Inserting $\frac{P_1}{\PTOT} = 1 - \frac{P_2}{\PTOT}$ and $\frac{P_2}{\PTOT}$ from Eq.~\ref{eq:p2_approx} into Eq.~\ref{eq:combinedflux} and Taylor expanding for small $\DELD_{12}$ gives
\begin{align}
\label{eq:flux_approx}
J_{12} + J_{21} \simeq e^{\TOT/2}\bigg[1 + \left(\frac{1}{2} - \frac{1}{1 + e^{-\TOT/2}}\right)(\DELD_{12})^2 + \frac{k_{\text{esc}}\DELD_{12}}{4(1 + e^{\TOT/2})}\bigg] - 1 \ .
\end{align}
We approximate the accumulated flux for $k_{\text{esc}}\frac{P_2}{\PTOT}t \ll 1$ as
\begin{subequations}
\begin{align}
\Phi_{\text{acc}} &= [J_{12}(t=0) + J_{21}(t=0)]\int_0^t e^{-k_{\text{esc}}\tfrac{P_2}{\PTOT}t'}dt' \\
\label{eq:acc_flux_approx}
&\simeq [J_{12}(t=0) + J_{21}(t=0)]\left(t - \frac{k_{\text{esc}}\frac{P_2}{\PTOT} t^2}{2}\right) \ ,
\end{align}
\end{subequations}
with the second line dropping the $t=0$ for the $J_{ij}$.
Inserting Eqs.~\ref{eq:p2_approx} and \ref{eq:flux_approx}, and their derivatives, into $\partial \Phi_{\text{acc}}/\partial\DELD_{12} = 0$ (using Eq.~\ref{eq:acc_flux_approx} for $\Phi_{\text{acc}}$), and dropping all terms of higher order than $k_{\text{esc}}$ and $\DELD_{12}$, produces Eq.~17 (reproduced here for convenience),
\begin{equation}
\label{eq:finalapproxone}
\frac{\Delta\DOPT_{12}}{k_{\text{esc}}} \simeq -\frac{1}{4}\left(t - \frac{1}{e^{\frac{1}{2}\TOT} - 1}\right) \ .
\end{equation}
A similar derivation for RL rate constants leads to Eq.~18,
\begin{equation}
\label{eq:finalapproxtwo}
\frac{\Delta\DOPT_{12}}{k_{\text{esc}}} \simeq -\frac{1}{4}\left(t + \frac{1}{1 - e^{-\frac{1}{2}\TOT}}\right) \ .
\end{equation}
We find the same equations using Maple 2016, keeping all non-Taylor expanded expressions, and then Taylor expanding the final derivative $\partial \Phi_{\text{acc}}/\partial\DELD_{12} = 0$ for small $k_{\text{esc}}$ and $\DELD_{12}$.

For both FL and RL, escape is reduced by a more negative 
$\Delta\DISS_{12}$.  
Reducing escape grows in importance for longer times, hence the negative sign of the $t$-dependent term in both Eqs.~\eqref{eq:finalapproxone} and \eqref{eq:finalapproxtwo}. 
For 
$\Delta\DISS_{12}=0$,
escape causes $P_2 < P_1$ at steady state, 
leading $[\partial(J_{12}+J_{21})/\partial\DISS_{12}]_{\Delta\DISS_{12}=0}$ to be positive for FL (increasing the forward rate constant from the state with larger steady-state probability) and negative for RL (increasing the reverse rate constant from the state with larger steady-state probability). 
Hence the $t$-independent term in $\Delta\DOPT_{12}$ is positive for FL (Eq.~\eqref{eq:finalapproxone})
and negative for RL (Eq.~\eqref{eq:finalapproxtwo}).
With FL and for large $\TOT$, the forward rate constants are very large, so escape influences $P_1$ and $P_2$ negligibly at $\Delta\DOPT_{12} = 0$, so the $t$-independent term of Eq.~\eqref{eq:finalapproxone}
decreases in magnitude towards zero as $\TOT$ increases.
With RL and for large $\TOT$, the reverse rate constants are very small, so the influence of escape on $P_1$ and $P_2$ becomes insensitive to $\TOT$ increases, so the $t$-independent term of Eq.~\eqref{eq:finalapproxtwo} decreases in magnitude to a constant value as $\TOT$ increases.

\begin{figure}[t] 
	\centering
	\hspace{-0.0in}
	\begin{tabular}{c}
		\hspace{-0.300in}\includegraphics[width=3.375in]{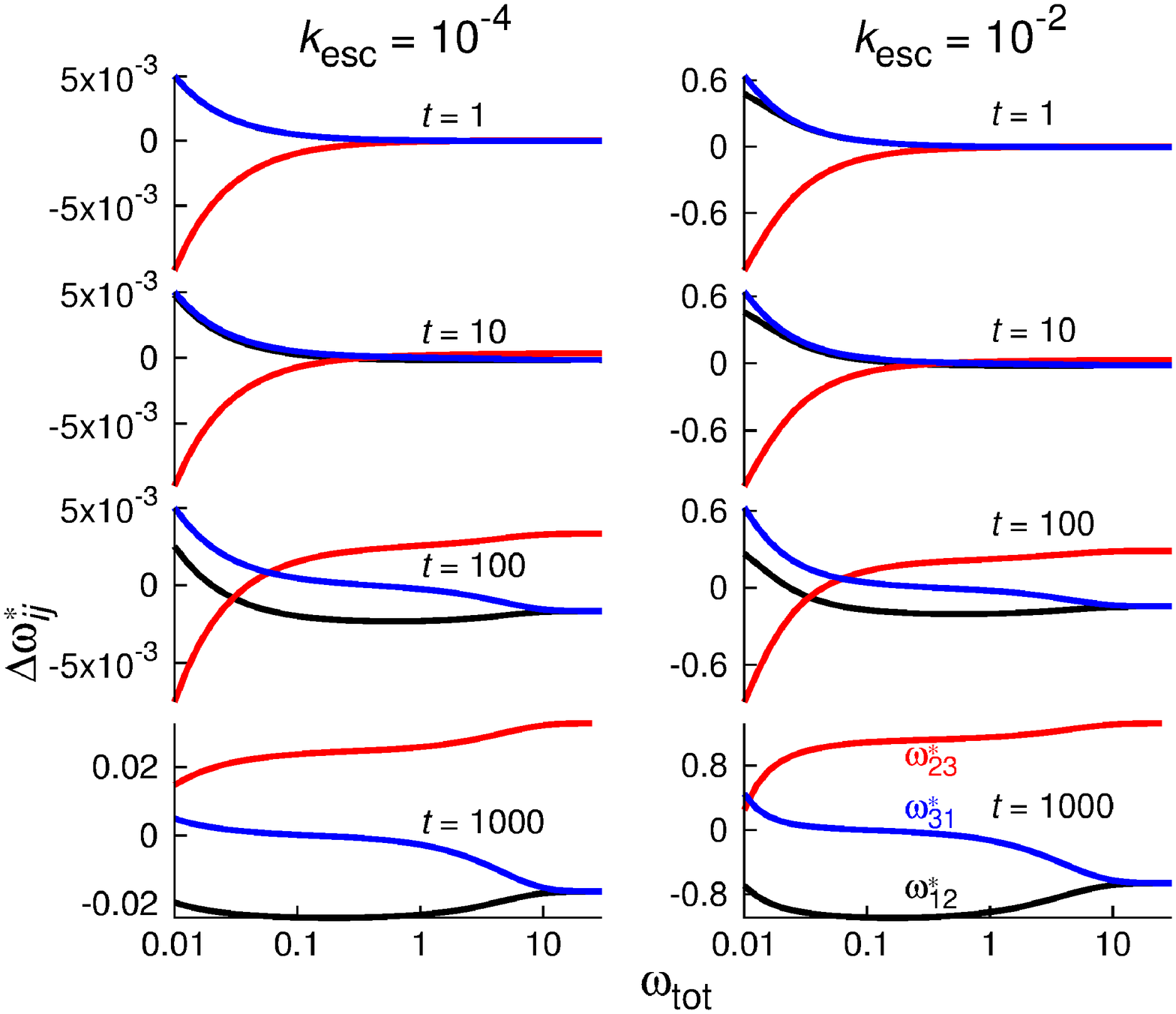}
	\end{tabular}
	\caption{\label{fig:ThreeStateEscape} 
{\bf Allocating free energy to maximize processivity of forward labile cycles.} 
Free energy allocation that maximizes accumulated flux up to time $t$ (Eq.~\ref{eq:AccumulatedFluxThreeState})
through three-state \forwardlabile cycle with state 2 vulnerable to escape. Bare rate constants $\BAREONETWO = \BARETWOTHREE = \BARETHREEONE = 1$.
}
\end{figure}

\begin{figure}[tbp] 
	\centering
	\hspace{-0.0in}
	\begin{tabular}{c}
		\hspace{-0.300in}\includegraphics[width=3.375in]{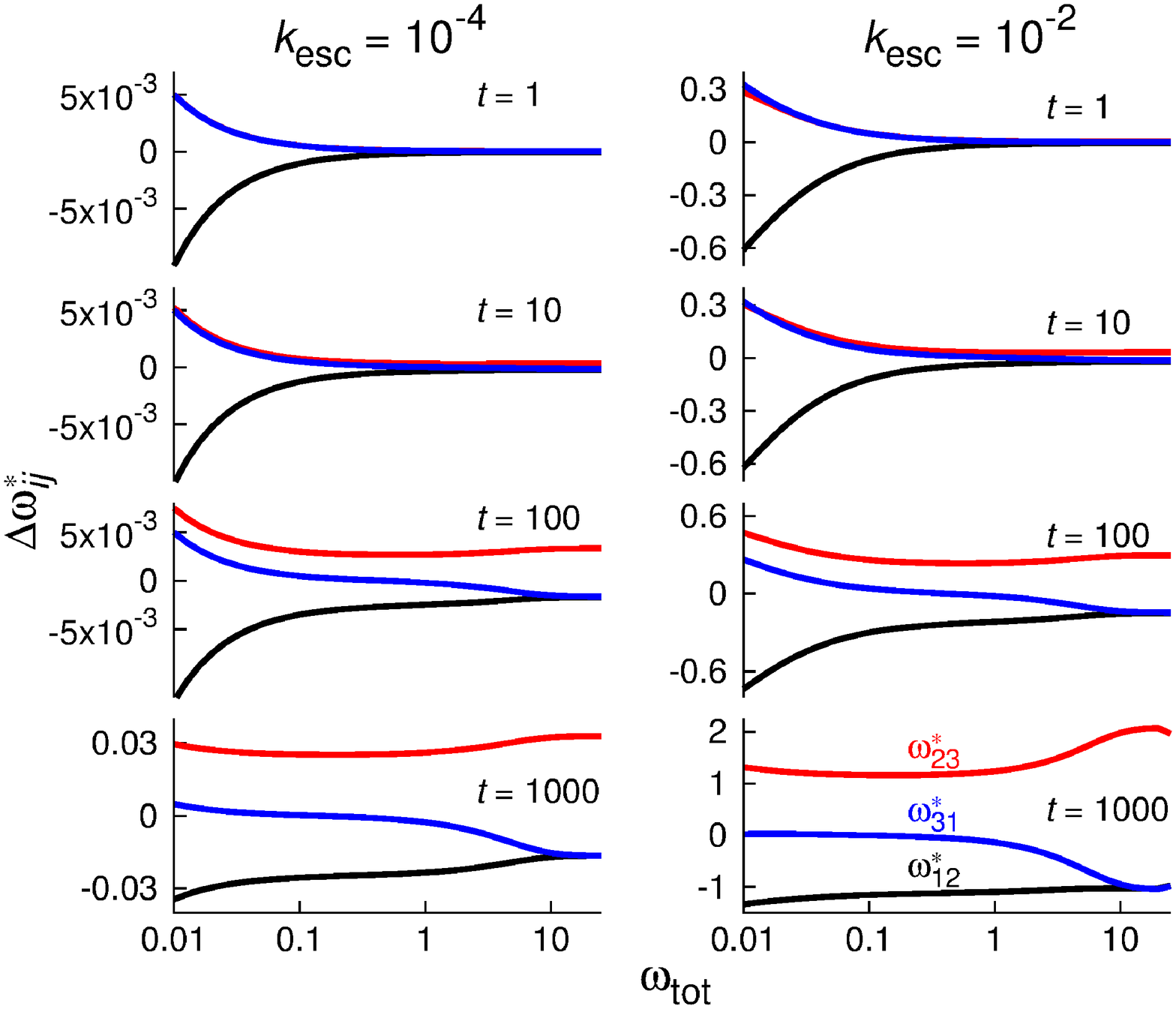}
	\end{tabular}
	\caption{\label{fig:ThreeStateEscapeBR} 
{\bf Allocating free energy to maximize processivity of reverse labile cycles.}
Free energy allocation
that maximizes accumulated flux up to time $t$ (Eq.~\ref{eq:AccumulatedFluxThreeState})
through three-state \reverselabile cycle with state 2 vulnerable to escape. Bare rate constants $\BAREONETWO = \BARETWOTHREE = \BARETHREEONE = 1$.
}
\end{figure}

\subsubsection{Three states}
The three-state cycle with state 2 vulnerable to escape is governed by differential equations
\begin{subequations}
\begin{align}
\label{eq:ShowEscapeThreeA}
\frac{\MD P_1}{\MD t} &= -(\RCFONETWO + \RCRTHREEONE)P_1 + \RCRONETWO P_2 + \RCFTHREEONE P_3 \\
\label{eq:ShowEscapeThreeB}
\frac{\MD P_2}{\MD t} &= \RCFONETWO P_1 - (\RCRONETWO + \RCFTWOTHREE + k_{\text{esc}})P_2 + \RCRTWOTHREE P_3 \\
\label{eq:ShowEscapeThreeC}
\frac{\MD P_3}{\MD t} &= \RCRTHREEONE P_1 + \RCRTWOTHREE P_2 - (\RCRTWOTHREE + \RCFTHREEONE)P_3 \ .
\end{align}
\end{subequations}

Substituting Eqs.~\ref{eq:ShowEscapeThreeA}, \ref{eq:ShowEscapeThreeB}, \ref{eq:ShowEscapeThreeC} into Eq.~\ref{eq:conversion} produces
\begin{subequations}
\begin{align}
\frac{\MD\left(\frac{P_1}{\PTOT}\right)}{\MD t} &= -(\RCFONETWO + \RCRTHREEONE)\frac{P_1}{\PTOT} + \RCRONETWO\frac{P_2}{\PTOT} + \RCFTHREEONE\frac{P_3}{\PTOT} + k_{\text{esc}}\frac{P_1}{\PTOT}\frac{P_2}{\PTOT} \\
\label{eq:ThreeStateConverted}
\frac{\MD \left(\frac{P_2}{\PTOT}\right)}{\MD t} &= \ \RCFONETWO\frac{P_1}{\PTOT} - (\RCRONETWO + \RCFTWOTHREE + k_{\text{esc}})\frac{P_2}{\PTOT} + \RCRTWOTHREE\frac{P_3}{\PTOT} + k_{\text{esc}}\left(\frac{P_2}{\PTOT}\right)^2 \\
\frac{\MD\left(\frac{P_3}{\PTOT}\right)}{\MD t} &= \ \RCRTHREEONE\frac{P_1}{\PTOT} + \RCFTWOTHREE\frac{P_2}{\PTOT} - (\RCRTWOTHREE + \RCFTHREEONE)\frac{P_3}{\PTOT} + k_{\text{esc}}\frac{P_3}{\PTOT}\frac{P_2}{\PTOT} \ .
\end{align}
\end{subequations}

Setting $\MD(P_i/\PTOT)/\MD t = 0$ gives
\begin{subequations}
\begin{align}
\frac{P_1}{\PTOT} &= \frac{\left(\RCFONETWO\RCRTWOTHREE + \RCFONETWO\RCFTHREEONE + \RCRTWOTHREE\RCRTHREEONE - \RCRONETWO k_{\text{esc}}\frac{P_2}{\PTOT}\right)\frac{P_2}{\PTOT}}{k_{\text{esc}}\left(k_{\text{esc}}\frac{P_2}{\PTOT} - \RCFONETWO - \RCRTWOTHREE - \RCFTHREEONE - \RCRTHREEONE\right)\frac{P_2}{\PTOT} + \RCFONETWO\RCRTWOTHREE + \RCFONETWO\RCFTHREEONE + \RCRTWOTHREE\RCRTHREEONE} \ , \\
\frac{P_3}{\PTOT} &= \frac{\frac{P_2}{\PTOT}}{\RCFTHREEONE}\left[\left(k_{\text{esc}}\frac{P_2}{\PTOT} - \RCFONETWO - \RCRTHREEONE\right)\xi - \RCRONETWO\right] \ , \\
\xi&\equiv
\dfrac{k_{\text{esc}}\RCRONETWO\frac{P_2}{\PTOT} - (\RCRONETWO\RCRTWOTHREE + \RCRONETWO\RCFTHREEONE + \RCFTWOTHREE\RCFTHREEONE)}
{k_{\text{esc}}^2\left(\frac{P_2}{\PTOT}\right)^2 - k_{\text{esc}}\left(\RCFONETWO + \RCRTWOTHREE + \RCFTHREEONE + \RCRTHREEONE\right)\frac{P_2}{\PTOT} + \RCFONETWO\RCRTWOTHREE + \RCFONETWO\RCFTHREEONE + \RCRTWOTHREE\RCRTHREEONE} \ .
\end{align}
\end{subequations}

Substituting these equations into Eq.~\ref{eq:ThreeStateConverted} produces a cubic equation for $P_2/\PTOT$,
\begin{align}
\label{eq:cubic}
&k_{\text{esc}}^2\left(\frac{P_2}{\PTOT}\right)^3 - k_{\text{esc}}(\RCFONETWO + \RCRONETWO + \RCFTWOTHREE + \RCRTWOTHREE + \RCFTHREEONE + \RCRTHREEONE + k_{\text{esc}})\left(\frac{P_2}{\PTOT}\right)^2\\
& + \big[k_{\text{esc}}(\RCFONETWO + \RCRTWOTHREE + \RCFTHREEONE + \RCRTHREEONE) + \RCFONETWO\RCFTWOTHREE + \RCFONETWO\RCRTWOTHREE + \RCFONETWO\RCFTHREEONE + \RCRTWOTHREE\RCRTHREEONE + \RCRONETWO\RCRTWOTHREE + \RCRONETWO\RCFTHREEONE + \RCRONETWO\RCRTHREEONE \nonumber\\
&+ \RCFTWOTHREE\RCFTHREEONE + \RCFTWOTHREE\RCRTHREEONE\big]\frac{P_2}{\PTOT}
 - (\RCFONETWO\RCRTWOTHREE + \RCFONETWO\RCFTHREEONE + \RCRTWOTHREE\RCRTHREEONE) = 0 \nonumber\ .
\end{align}
Cubic equations 
generally have at least one real solution, so a real $P_2/\PTOT$ exists.

Solving the cubic for $P_2/\PTOT$ in Eq.~\ref{eq:cubic} determines the steady-state $P_i/\PTOT$ and $\PTOT(t)$.
Because the flux decays with time as probability escapes, we additionally consider the rate of escape to evaluate molecular machine progress. Instead of flux alone, we combine flux and escape into the accumulated flux 
\begin{equation}
\label{eq:AccumulatedFluxThreeState}
\Phi_{\text{acc}} = \int_0^t\left[J_{12}(t') + J_{23}(t') + J_{31}(t')\right] \MD t' \ ,
\end{equation}
which we maximize after a time $t$. 
Figs.~\ref{fig:ThreeStateEscape} (FL cycles) and \ref{fig:ThreeStateEscapeBR} (RL cycles) show $\DISS_{ij}$ allocations which maximize $\Phi_{\text{acc}}$. These three-state results are similar to the two-state results in the main text.

\let\newpage = \oldnewpage

\bibliography{BrownSivak}

\end{document}